# Evaluating Multi-Agent LLM Architectures for Rare Disease Diagnosis

Ahmed Almasoud

*Artificial Intelligence and Data Analytics (AIDA) Lab, College of Computing and Information Sciences, Prince Sultan University, Riyadh, Saudi Arabia.*

Email: almasoud@psu.edu.sa

## Abstract

While large language models are capable diagnostic tools, the impact of multi-agent topology on diagnostic accuracy remains underexplored. This study evaluates four agent topologies, Control (single agent), Hierarchical, Adversarial, and Collaborative, across 302 cases spanning 33 rare disease categories. We introduce a Reasoning Gap metric to quantify the difference between internal knowledge retrieval and final diagnostic accuracy. Results indicate that the Hierarchical topology (50.0% accuracy) marginally outperforms Collaborative (49.8%) and Control (48.5%) configurations. In contrast, the Adversarial model significantly degrades performance (27.3%), exhibiting a massive Reasoning Gap where valid diagnoses were rejected due to artificial doubt. Across all architectures, performance was strongest in Allergic diseases and Toxic Effects categories but poorest in Cardiac Malformation and Respiratory cases. Critically, while the single-agent baseline was generally robust, all multi-agent systems, including the Adversarial model, yielded superior accuracy in Bone and Thoracic disease categories. These findings demonstrate that increasing system complexity does not guarantee better reasoning, supporting a shift toward dynamic topology selection.

## 1. Introduction

The integration of Large Language Models (LLMs) into clinical workflows represents a major shift in diagnostic assistance. Recent meta-analyses indicate that LLM assistance can improve the diagnostic accuracy of physicians across various medical fields and career stages [1]. However, the application of these models to rare and complex pathologies remains fraught with challenges. Benchmarking studies reveal that while generalist models perform well on common presentations, they frequently lack the precision of traditional bioinformatics tools when confronting rare genetic diseases [2], with some state-of-the-art models achieving accuracies as low as 16.5% on narrative-based rare disease tasks [3].

To mitigate the hallucinations and reasoning failures inherent in single-model inference, researchers are increasingly turning toward Multi-Agent Systems (MAS). These architectures attempt to replicate clinical Communities of Practice, such as Multi-Disciplinary Teams (MDTs) or tiered supervisory structures [4]. Existing research, such as MedChat [5] and AI Hospital [6], demonstrate that distributing tasks across role-specific agents can enhance reliability and simulation fidelity [7]. Furthermore, theoretical advancements in the general domain suggest that multi-agent debate can improve factuality and reduce errors through social correction [8].

Despite this progress, the impact of specific agent topologies on diagnostic precision remains underexplored. This study addresses this gap by evaluating four distinct agent architectures, Control, Hierarchical, Adversarial, and Collaborative, across 302 rare disease cases curated by Chen et al. [9]. We also introduce a novel metric, the *Reasoning Gap*, to quantify the divergence between internal knowledge retrieval and final adjudication.

## 2. Related Work

The literature surrounding AI-driven medical diagnosis has evolved from evaluating isolated LLMs to designing complex, agentic workflows. Dinc et al. [10] observed that while advanced models like Claude 3.7 achieve near-perfect accuracy on common cases, performance varies significantly in complex scenarios. This limitation is particularly high in rare diseases; Reese et al. [2] demonstrated that generalist LLMs fail to match the diagnostic ranking performance of phenotype-specific tools like Exomiser. Similarly, Gupta et al. [3] used a dataset derived from House M.D. to show that even frontier models struggle with the long tail of medical anomalies, necessitating architectures that go beyond simple zero-shot prompting.

### 2.1 Hierarchical and Role-Based Architectures

To address single-agent deficiencies, recent research emphasizes Agentic Oversight and hierarchical delegation. Kim et al. [11] introduced Tiered Agentic Oversight (TAO), mimicking hospital hierarchies (e.g., nurse-physician-specialist). The findings indicate that adaptive, layered supervision can absorb up to 24% of individual errors, suggesting that structural hierarchy acts as a noise filter. This aligns with the work of Zuo et al. [12], whose KG4Diagnosis framework employs a General Practitioner agent for triage and specialist agents for in-depth analysis, enhanced by Knowledge Graphs.

The utility of role-playing agents is further supported by Wang, Z. et al. [13], who proposed MedAgent-Pro. By decoupling diagnosis into disease-level planning and patient-level reasoning, the research demonstrated that structured, step-by-step workflows reduce empirical hallucinations, a finding reinforced by Peng et al. [14], who implemented a tree-of-reasoning framework. Specific clinical domains have also benefited from this approach; for instance, Ji et al. [15] developed SpAgents for axial spondyloarthritis, where distinct planner, data, and doctor agents collaborated to outperform junior rheumatologists. Similarly, researchers utilized hybrid agents to emulate tumor boards in oncology [16] and cancer care [17], confirming that multi-modal, multi-perspective systems align better with clinical consensus than monolithic models.

## 2.2 Debate, Collaboration, and Aggregation Mechanisms

A critical divergence in multi-agent design is the method of consensus. Chen et al. [9] demonstrated that a Multi-Agent Conversation (MAC) framework, inspired by MDT discussions, outperformed Chain-of-Thought (CoT) and Self-Consistency methods in rare disease diagnosis. This collaborative approach is reinforced by Thakrar et al. [18], who found that clinical-inspired reasoning layers, which combine perspectives analogous to peer consultation, yielded higher accuracy than fine-tuning alone.

Meanwhile, Du et al. [8] argued that forcing agents to critique and debate responses enhances factuality and strategic reasoning in general domains. However, the translation of this debate mechanism to medicine is contested. While Feng, Y. et al. [19] utilized Reinforcement Learning to optimize questioning strategies in multi-turn dialogues, the effectiveness of pure debate in resolving medical ambiguity remains an open question. Furthermore, the aggregation of these multi-agent outputs presents its own challenges. Ai et al. [20] proposed algorithms like "Inverse Surprising Popularity" to mitigate the limitations of simple majority voting, suggesting that the correlation and heterogeneity of agents must be mathematically accounted for to ensure reliable collective decisions.

Our research builds upon these foundational studies by directly comparing the collaborative and hierarchical models favored in clinical literature [9, 11] against the adversarial debate models proposed in general AI research [8], specifically within the context of rare disease diagnosis.

## 3. Methodology

Figure 1a provides a schematic overview of the research methodology. All diagnostic experiments were conducted using the GPT-5.1 large language model accessed via OpenAI API.

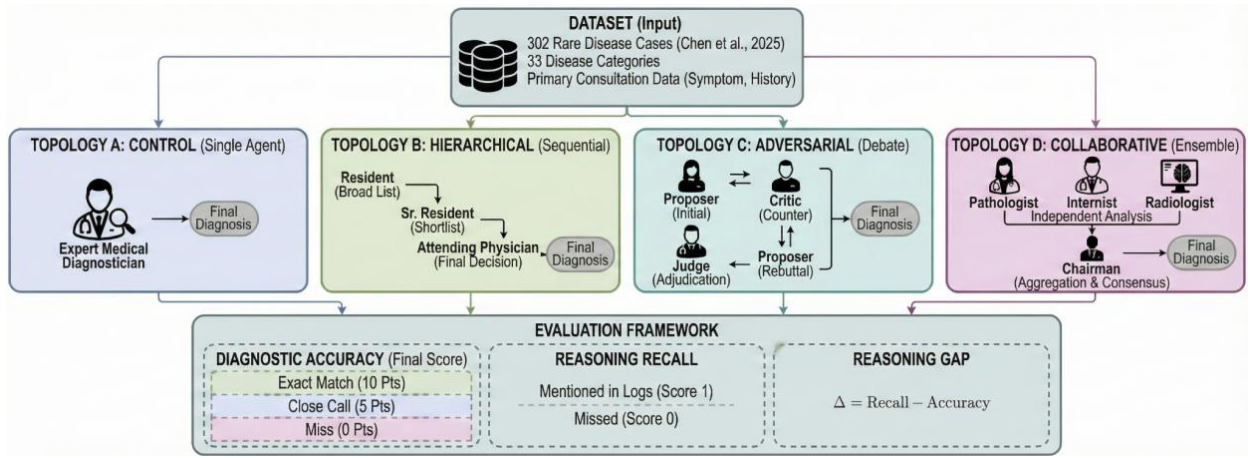

Figure 1a. Schematic overview of the experimental methodology.

## 3.1 Dataset

This study uses a publicly available dataset of 302 rare disease cases curated byc Chen et al. [9]. The cases were derived from real-world clinical reports, covering 33 distinct disease categories sourced from the Orphanet Database. Each entry represents a Primary Consultation scenario, providing patient demographics, physical examination, medical history, and initial test results to replicate the ambiguity of a first clinical encounter. The dataset includes the confirmed final diagnosis from the original case report, which serves as the ground truth for our evaluation.

## 3.2 Multi-Agent Architectures

To evaluate the impact of different reasoning structures on diagnostic accuracy, we designed four distinct topological configurations. These range from a standard single-agent baseline to complex multi-agent systems that mimic varying clinical workflows.

### 3.2.1 Control (Single Agent)

The control topology represents the baseline performance of the LLM. It uses a direct, zero-shot prompting strategy where a single agent, acting as an "Expert Medical Diagnostician," analyzes the patient data and outputs the single most likely diagnosis without intermediate reasoning steps or external feedback. This serves as the benchmark against which the multi-agent architectures are compared.

### 3.2.2 Hierarchical (Sequential)

This topology mimics a tiered hospital workflow designed to filter a broad differential diagnosis down to a specific conclusion. As detailed in **Protocol 1.1** (Figure 1b), we structured the system as a three-stage funnel. The process begins with a **Resident** agent who analyzes the case and generates a list of three potential diagnoses. This output is then passed to a **Senior Resident**, who reviews the Resident's list, rules out an unlikely candidate based on symptom analysis, and shortlists the top two diagnoses. Then, an **Attending Physician** makes the final decision by selecting the single most likely diagnosis from the Senior Resident's shortlist.

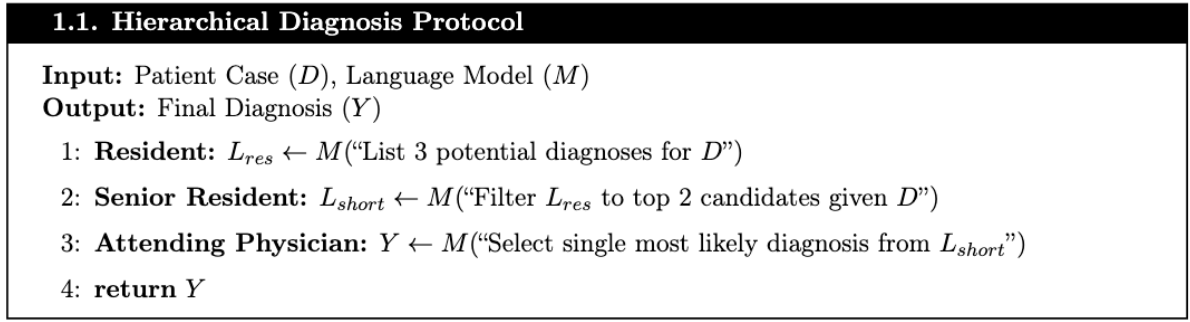

Figure 1b. Hierarchical Diagnosis Protocol

### 3.2.3 Adversarial (Debate)

This topology introduces conflict-driven reasoning to mitigate confirmation bias. As outlined in **Protocol 1.2** (Figure 1c), the system forces a structured debate between a **Proposer** and a **Critic**. Unlike the sequential model, this topology explicitly instructs the Critic agent to find contradictory evidence regardless of the initial diagnosis's strength. This ensures that the final adjudicator ("The Judge") is presented with a stress-tested argument rather than a single uncontested viewpoint. The judge then reviews the entire exchange (Proposal, Critique, Rebuttal) to adjudicate the final diagnosis.

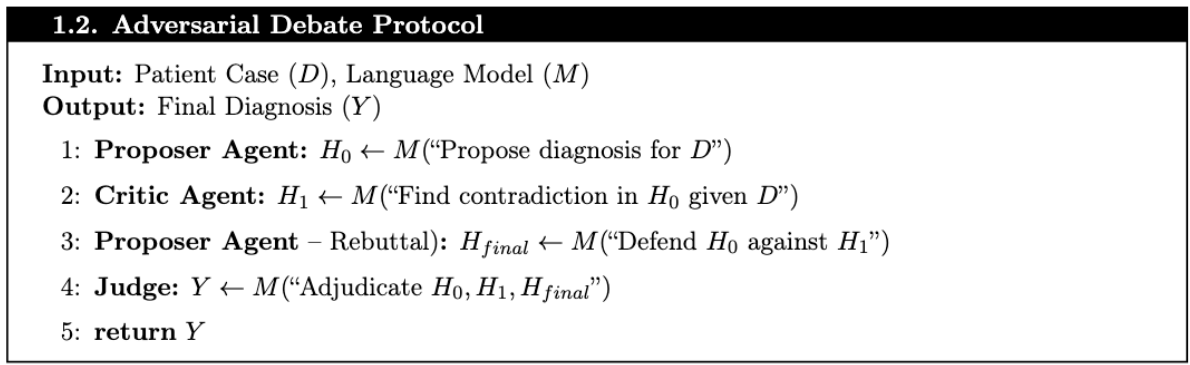

Figure 1c. Adversarial Debate Protocol

### 3.2.4 Collaborative (Ensemble)

This topology leverages the wisdom of the crowd by simulating a Multi-Disciplinary Team (MDT) approach. As shown in **Protocol 1.3** (Figure 1d), three specialist agents, a Pathologist, Internist, and Radiologist, analyze the case simultaneously and independently. Then, the Chairman agent reviews the case data alongside the three independent opinions to determine a final consensus diagnosis.

```
1.3. Collaborative Ensemble Protocol
Input: Patient Case (D), Language Model (M)
Output: Final Diagnosis (Y)
 1: Parallel Analysis:
 2:    O_path ← M("Pathologist: Analyze D")
 3:    O_int  ← M("Internist: Analyze D")
 4:    O_rad  ← M("Radiologist: Analyze D")
 5: Aggregation:
 6:    Chairman: Y ← M("Synthesize {O_path, O_int, O_rad} into consensus")
 7: return Y
```

Figure 1d. Collaborative Ensemble Protocol

## 3.3 Evaluation Framework

To rigorously assess the performance of each topology, we implemented an evaluation system designed to measure both the accuracy of the final output and the breadth of the underlying reasoning process.

### 3.3.1 Diagnostic Accuracy Score

We used LLM-as-a-Judge approach to evaluate the models, with GPT-5.1 assigning a score to each diagnosis. The automated adjudicator (LLM) compared the model's prediction against the ground truth using a three-tier scoring rubric that distinguishes between exact matches (10), clinically relevant differentials (5), and complete misses (0). The full rubric logic is provided in **Figure 2**.

To allow for comparative analysis, the final **Diagnostic Accuracy Score** was calculated as the mean score across the dataset, normalized to a percentage scale:

$$\text{Diagnostic Accuracy Score} = \left( \frac{1}{N} \sum_{i=1}^{N} S_i \right) \times 10$$

where N=302 represents the total number of cases and $S_i \in \{0, 5, 10\}$ represents the score assigned to case i. By incorporating partial credit for relevant differentials, this metric reflects diagnostic utility beyond strict binary correctness.

While recent studies suggest LLMs may exhibit self-preference bias in open-ended evaluations [21], our evaluation directly mitigated this by using ground-truth and a scoring rubric (Figure 2). This constrains the evaluator to objective diagnostic criteria rather than subjective preferences.

Furthermore, since the same judge (GPT-5.1) evaluated all four topologies (Control, Hierarchical, Adversarial, Collaborative), any residual bias remains constant across experimental conditions, ensuring the validity of the comparative analysis.

> **Evaluation Rubric (LLM-as-a-Judge)**
>
> **System Instruction:** You are a Medical Adjudicator. Compare the Prediction vs Ground Truth.
> **SCORING RULES:**
> **[10 POINTS] Exact Match**
> The diagnosis is identical to the ground truth or is a recognized direct synonym (e.g., "Stein-Leventhal Syndrome" == "PCOS"). Ignore minor formatting differences.
>
> **[5 POINTS] Close Call / Differential**
> The predicted disease is in the exact same family or is a direct subtype/parent of the truth.
> *Example:* Truth="Duchenne Muscular Dystrophy", Prediction="Muscular Dystrophy" (Generic).
> *Example:* Truth="Viral Meningitis", Prediction="Meningitis" (Unspecified).
>
> **[0 POINTS] Miss**
> The prediction involves a completely unrelated pathology, organ system, or etiology.

Figure 2. Evaluation Rubric

### 3.3.2 Reasoning Gap

To quantify the efficiency of the decision-making process, we introduced a novel metric termed the **Reasoning Gap**. This metric distinguishes between *retrieval failure* (the model never identifying the correct pathology) and *adjudication failure* (the model rejecting the correct diagnosis during deliberation).

First, we calculated a baseline **Reasoning Recall** score by scanning the full interaction logs (e.g., the Resident's initial list or the Proposer's arguments) to determine if the correct diagnosis was ever explicitly considered, regardless of the final output. This involves assigning a score of 10 if the ground truth diagnosis was explicitly present (via **strict keyword matching**), and 0 if it was absent. These binary scores were then averaged and normalized to a percentage scale (0–100%) for each topology.

We defined the Reasoning Gap as the difference between this potential (Recall) and the realized performance (Accuracy):

$$\Delta = \text{Reasoning Recall } (\%) - \text{Diagnostic Accuracy } (\%)$$

This derived metric serves as the Rejection Error. A large positive gap indicates that the topology has the correct diagnosis in its knowledge base but failed to select it as the final answer e.g., due to persuasive interference (in Adversarial models) or excessive filtering (in Hierarchical models). Conversely, a minimal gap indicates a highly efficient decision boundary where recognized patterns are correctly converted into final diagnoses.

### 3.3.3 Subgroup Analysis

To evaluate the robustness of the topologies across diverse medical domains, we conducted a stratified analysis based on the **Disease Type**. Performance scores were aggregated across 33 distinct clinical categories (e.g., *Neoplastic, Infectious, Genetic, Immunological*) to identify specific domain strengths or weaknesses inherent to each multi-agent architecture.

## 4. Results and Discussion

### 4.1 Overall Diagnostic Performance

The comparative performance of the four topological architectures reveals a non-linear relationship between system complexity and diagnostic success. As summarized in **Table 1** and visualized in **Figure 3**, the structural arrangement of agents influenced both the retrieval of correct medical knowledge (Reasoning Recall) and the ability to finalize that knowledge into a correct diagnosis (Accuracy).

**Table 1. Summary of Performance Metrics across Topologies**

| Topology | Diagnostic Accuracy (%) | Reasoning Recall (%) | Reasoning Gap (Δ) |
| --- | --- | --- | --- |
| **Control (Baseline)** | 48.5% | N/A | N/A |
| **Hierarchical** | **50.0%** | **54.0%** | 4.0 |
| **Collaborative** | 49.8% | 51.3% | 1.5 |
| **Adversarial** | 27.3% | 44.0% | **16.7** |

The **Hierarchical topology** emerged as the best-performing architecture, achieving a diagnostic accuracy of **50.0%**. This represents a modest but consistent improvement (+1.5%) over the single-agent **Control** baseline (48.5%). Notably, the Hierarchical model also demonstrated the highest **Reasoning Recall (54.0%)**, suggesting that the sequential funnel structure, starting with a broad differential by the Resident agent and refining it through senior review was the most effective strategy for surfacing the correct ground truth. This is consistent with findings on Tiered Agentic Oversight, where layered supervision was shown to effectively absorb individual agent errors **(Kim et al., 2025).**

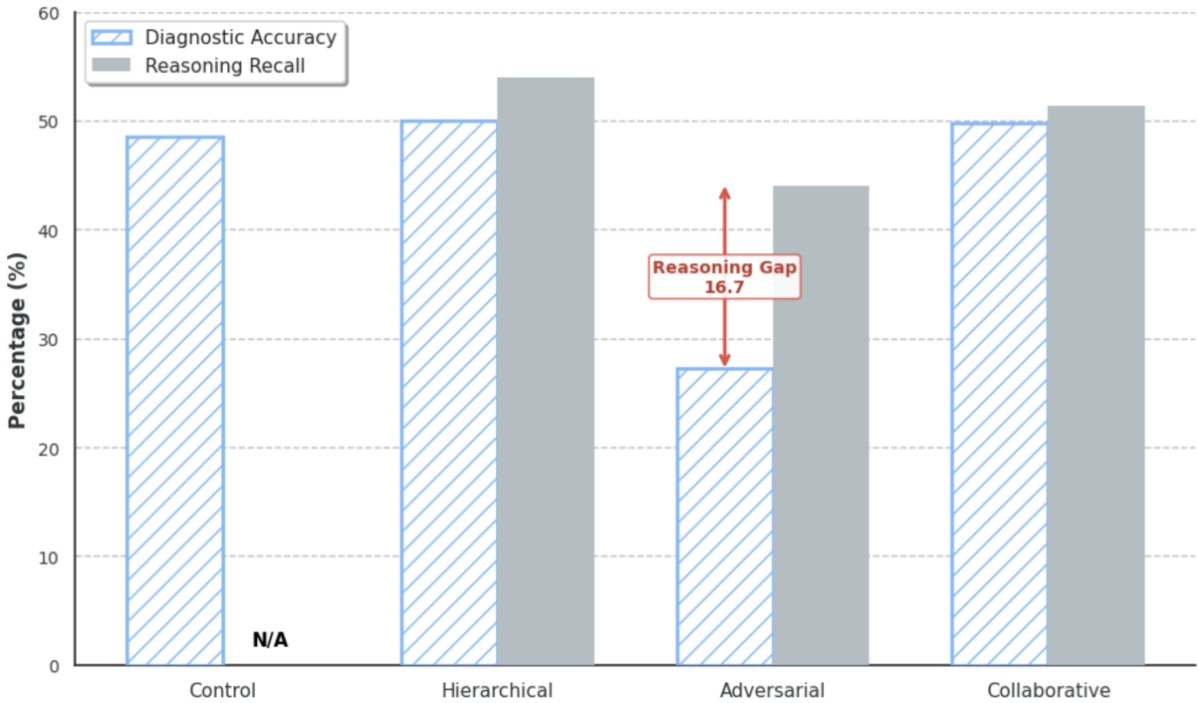

Figure 3. Performance across topologies

*Note: Reasoning Recall is marked as N/A for the Control topology because it directly outputs only the final diagnosis without the intermediate interaction logs generated by the multi-agent topologies.*

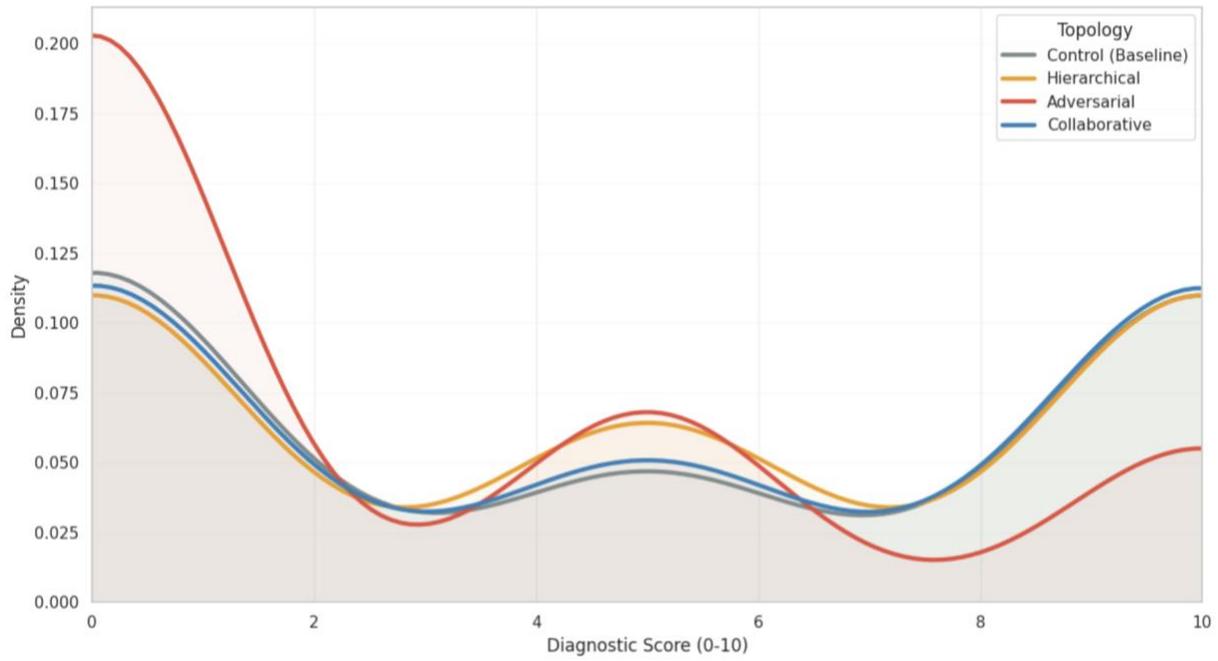

Figure 4. Probability density of diagnostic scores across topologies (multi-agents) and control (single agent)

The **Collaborative topology** (Ensemble) performed comparably to the Hierarchical model in final output, achieving a diagnostic accuracy of **49.8%** and a recall of **51.3%**. The most striking finding, however, was the substantial performance degradation observed in the **Adversarial topology**. Despite achieving a reasoning recall of **44.0%**, the final diagnostic accuracy plummeted to **27.3%**. As illustrated in the probability density plot (**Figure 4**), the Adversarial model's score distribution shifts dramatically toward the lower end of the spectrum. This indicates a systemic failure in the decision boundary: while the Proposer or Critic agents often identified the correct disease during the debate, the Judge agent was frequently swayed by counter-arguments, leading to the rejection of valid diagnoses in favor of plausible but incorrect alternatives.

These results challenge the assumption that increasing agent interaction inherently improves reasoning. While the addition of agents in the Hierarchical and Collaborative models improved the system's *performance marginally* over the single agent, the introduction of conflict in the Adversarial model introduced destructive interference, severely hampering the system's *precision*.

## 4.2. Analysis of the Reasoning Gap

We analyzed the **Reasoning Gap** (Δ), defined as the quantitative difference between **Reasoning Recall** (knowledge retrieval) and **Diagnostic Accuracy** (final decision correctness). The Hierarchical model maintained a moderate gap of 4.0, suggesting that its multi-stage review process (Resident → Senior → Attending) functioned as an effective clinical filter. Similarly, the Collaborative topology exhibited a low gap of 1.5, indicating that the Chairman agent rarely overruled the specialist consensus.

The most significant anomaly was observed in the Adversarial topology, which exhibited a massive Reasoning Gap of 16.7. This gap indicates a systemic breakdown in the adjudication phase. Qualitative analysis of the interaction logs suggests that the constraint of mandatory critique frequently obscured the correct diagnosis. Given that the Critic agent was instructed to find contradictory evidence regardless of the initial diagnosis's strength, it often generated plausible but ultimately irrelevant distractors. In instances where the Proposer correctly identified the disease, the Critic's forced skepticism introduced artificial doubt; conversely, when the Critic correctly identified a missed diagnosis, the Proposer's defensive rebuttal often persuaded the Judge to adhere to the incorrect initial hypothesis. This indicates that while debate increases idea generation, unconditional skepticism can degrade the precision required for rare disease diagnosis.

## 4.3 Domain-Specific Performance (Subgroup Analysis)

We analyzed the performance across 33 disease categories as shown in Table 2. This analysis, visualized in the Performance Gain vs. Control (Figure 6) and the Diagnostic Difficulty (Figure 5), highlights distinct behavioral patterns driven by disease complexity.

Table 2. Breakdown of each topology diagnostic scores by disease category based on the evaluation rubric

| Disease Type | Control | Hierarchical | Adversarial | Collaborative |
|---|---|---|---|---|
| **Abdominal Surgical** | 2.50 | 2.50 | 1.25 | 2.50 |
| **Allergic** | 10.00 | 9.00 | 4.00 | 10.00 |
| **Bone** | 3.85 | 5.77 | 4.62 | 5.38 |
| **Cardiac** | 6.25 | 6.25 | 3.75 | 6.25 |
| **Cardiac Malformation** | 1.25 | 1.25 | 1.25 | 1.25 |
| **Circulatory** | 6.25 | 5.63 | 4.38 | 5.00 |
| **Embryogenesis** | 2.14 | 3.57 | 1.43 | 3.57 |
| **Endocrine** | 5.50 | 7.00 | 4.00 | 6.00 |
| **Gastroenterological** | 6.25 | 6.25 | 1.88 | 8.13 |
| **Genetic** | 3.67 | 4.67 | 2.33 | 3.33 |
| **Gynae. & Obstetric** | 5.56 | 4.44 | 2.22 | 5.56 |
| **Haematological** | 5.00 | 5.00 | 5.00 | 4.00 |
| **Hepatic** | 8.75 | 7.50 | 3.13 | 8.75 |
| **Immunological** | 5.00 | 5.50 | 2.50 | 4.00 |
| **Inborn Metabolism** | 3.64 | 3.64 | 2.73 | 3.64 |
| **Infectious** | 3.75 | 5.63 | 1.25 | 5.00 |
| **Infertility** | 5.63 | 5.63 | 1.88 | 5.63 |
| **Neoplastic** | 4.17 | 2.50 | 1.67 | 3.33 |
| **Neurological** | 4.23 | 4.23 | 2.69 | 4.62 |
| **Odontological** | 5.00 | 7.14 | 3.57 | 6.43 |
| **Ophthalmic** | 5.00 | 4.09 | 2.27 | 4.55 |
| **Otorhinolaryngological** | 6.00 | 5.00 | 3.00 | 6.00 |
| **Renal** | 3.33 | 3.33 | 2.78 | 3.89 |
| **Respiratory** | 1.43 | 0.71 | 1.43 | 5.00 |
| **Rheumatological** | 5.00 | 6.00 | 3.00 | 5.50 |
| **Rheum. (Childhood)** | 10.00 | 8.33 | 5.00 | 8.33 |
| **Skin** | 2.73 | 5.00 | 1.82 | 3.18 |
| **Surgical Maxillo-facial** | 3.33 | 5.00 | 2.50 | 4.17 |
| **Teratologic** | 6.67 | 5.83 | 2.50 | 5.00 |
| **Thoracic Surgical** | 2.50 | 5.00 | 4.38 | 3.13 |
| **Toxic Effects** | 8.33 | 8.33 | 3.33 | 8.33 |
| **Transplant Related** | 5.91 | 4.09 | 3.18 | 4.55 |
| **Urogenital** | 5.63 | 7.50 | 3.13 | 8.13 |

The intrinsic difficulty of cases varied significantly. As shown in Figure 5, categories such as Allergic Diseases (9.7 - avg. score, excluding adversarial topology), Toxic Effects (8.3), and Hepatic Diseases (8.3) proved to be highly solvable for the LLM. Conversely, categories with high structural ambiguity, such as Cardiac Malformations (1.2) and Abdominal Surgical cases (2.5), remained resistant to diagnosis regardless of the architectural topology. However, a notable divergence was observed in respiratory diseases. While the Control and Hierarchical models struggled significantly with these cases (scoring 1.43 and 0.71, respectively), the Collaborative topology achieved a score of 5.00. This improvement suggests that the multi-perspective ensemble was uniquely capable of synthesizing the complex, overlapping symptoms inherent to respiratory pathology.

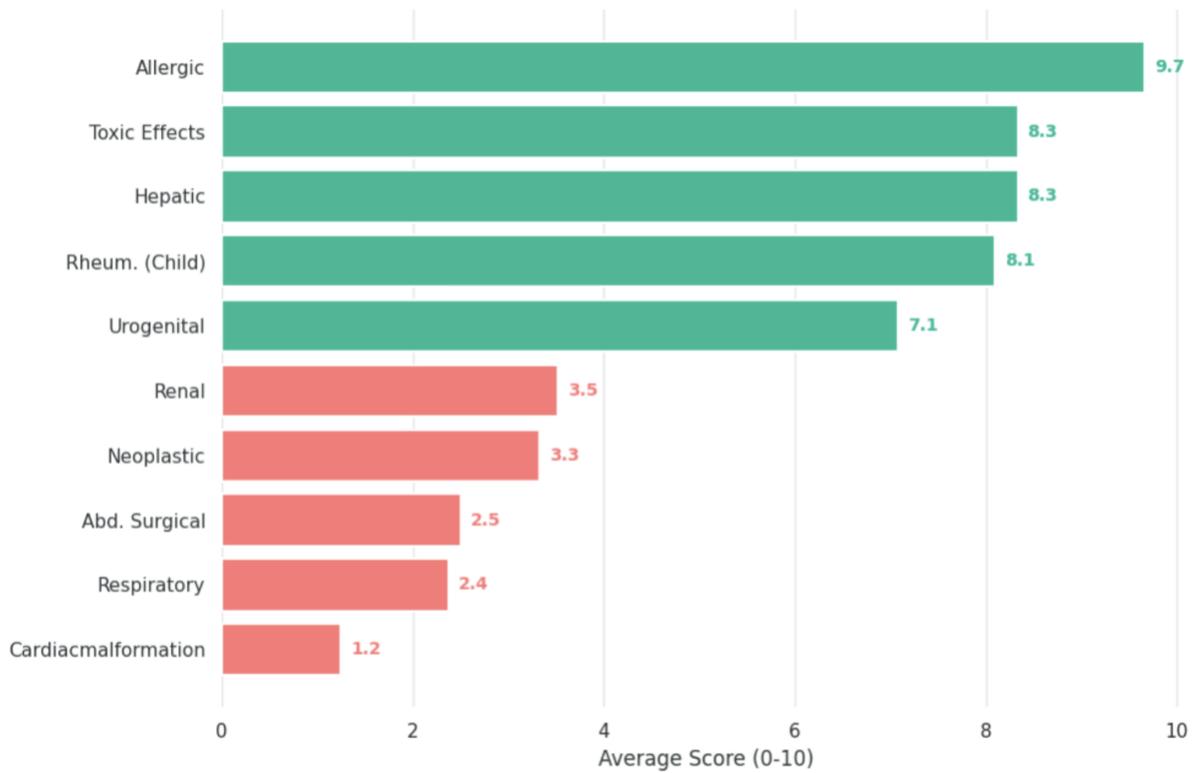

Figure 5. Average diagnostic performance across disease categories (excluding Adversarial topology)

*Note*: Scores represent the mean Diagnostic Accuracy Score (0-10) aggregated across Control, Hierarchical, and Collaborative topologies. The Adversarial topology was excluded from this visualization to highlight intrinsic disease difficulty without the destructive interference observed in debate-based models.

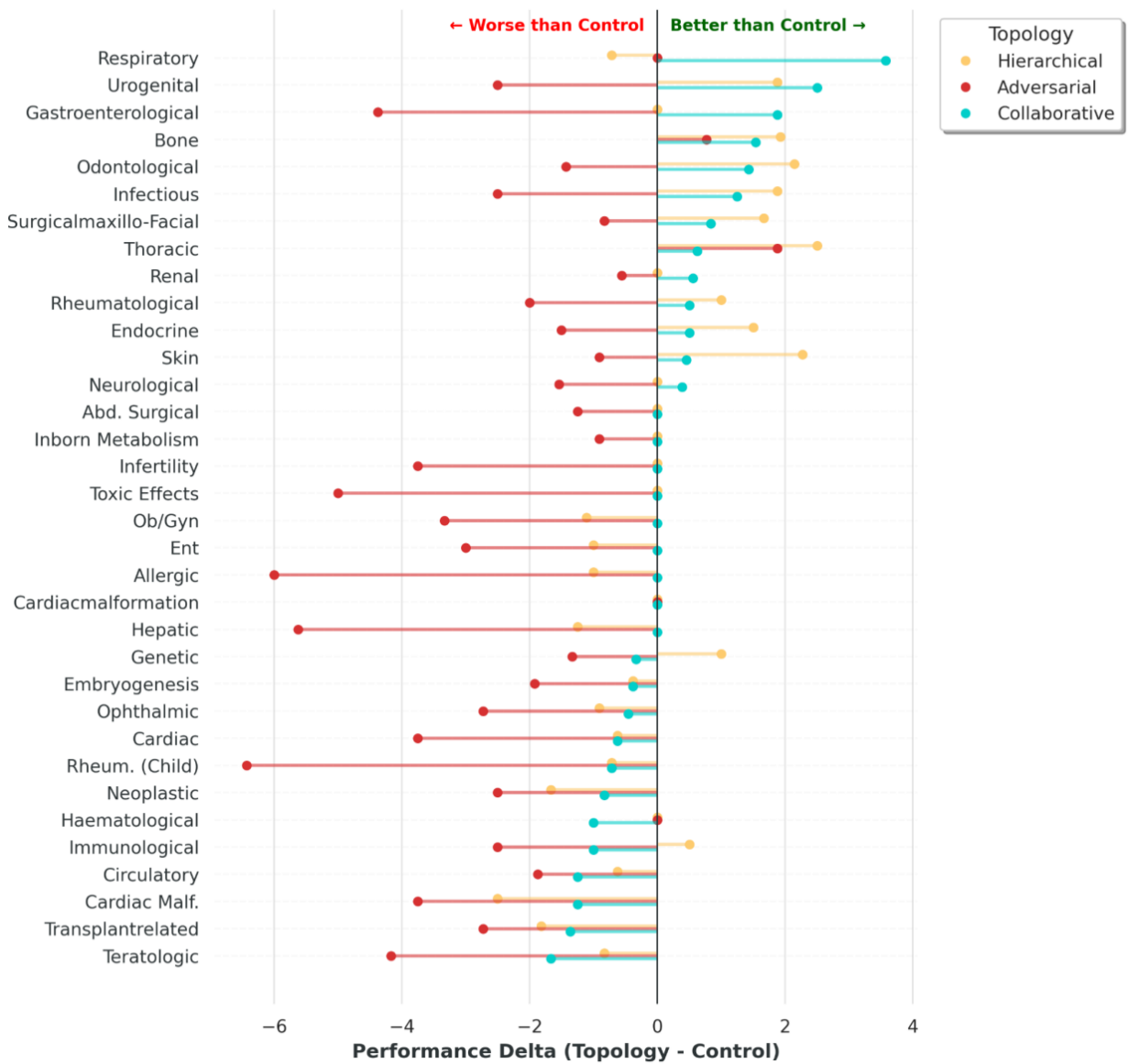

Figure 6. Performance of each topology vs control (baseline)

Figure 6 reveals how each multi-agent architecture performed relative to the Control baseline (represented by the central axis at x=0). A striking observation is the robustness of the single-agent Control. In some complex categories, specifically Teratologic, Transplant-related, and Circulatory, the Control agent outperformed the multi-agent ensembles, which appear to the left of the baseline. This challenges the assumption that adding more agents always improves results. The Collaborative topology did show strong gains in cases (involving overlapping organ systems) like Respiratory (+3.6) and Urogenital diseases, where synthesizing different symptoms is key. However, its failure to beat the single agent in some categories raises valid questions about efficiency.

In sharp contrast, the Adversarial topology consistently appears to the left of the baseline, indicating systemic performance degradation. Crucially, this collapse was most severe in the easiest diagnostic categories. In Allergic Diseases and Rheumatological (Child) cases, where the Control model scored near-perfectly, the Adversarial agent exhibited substantial negative deltas of approximately -6.0 points. This phenomenon confirms the Rejection Error hypothesis proposed in Section 4.2: when the case is unambiguous, the introduction of a Critic creates artificial doubt. This triggers the adjudicator to overthink the evidence, ultimately rejecting a clear, correct solution in favour of a plausible distractor generated during the debate.

### 4.4 Limitations

While this study provides a comprehensive evaluation of multi-agent topologies, several limitations must be acknowledged. First, our reliance on a specific large language model (GPT-5.1) introduces a model-dependent bias. Future work should replicate this setup across diverse model families (e.g., Claude, Llama, Gemini).

Second, the dataset was restricted to 302 Primary Consultation cases derived from the Orphanet database. As these cases focus specifically on rare diseases, the findings may not generalize perfectly to common clinical presentations where symptom overlap is more frequent. In addition, this study treated the diagnostic process as a static, single-turn inference task. In clinical reality, diagnosis is an iterative process involving inquiry, lab testing, and patient feedback [22]. Future iterations of this work should extend our topological evaluation into dynamic environments by leveraging multi-turn dialogue frameworks such as DoctorAgent-RL [19] and MedAgentSim [23].

Finally, we must consider the cost-benefit trade-off. The Hierarchical and Collaborative models require significantly higher token usage and inference time compared to the Control. Given that the diagnostic accuracy gain was marginal (<2%), the computational overhead of complex multi-agent systems may not be justifiable for general screening, though it may hold value for specific, high-stakes diagnostic sub-categories.

## 5. Conclusion

This study provides a systematic evaluation of multi-agent systems in clinical reasoning, offering a detailed analysis of how architectural design impacts diagnostic precision. Our key contributions are as follows:

- We provided an empirical comparison of four distinct agent architectures, Control, Hierarchical, Adversarial, and Collaborative, across 302 rare disease cases. Our results identify the **Hierarchical** as the best performing topology (50.0% accuracy), closely followed by Collaborative and Control.

- We introduced a novel metric, Reasoning Gap (Δ = Reasoning Recall - Diagnostic Accuracy) to measure the distance between a lack of knowledge (retrieval failure) and a failure of judgment (adjudication failure).
- We demonstrated a critical failure mode in the Adversarial topology which suffered a massive 16.7-point Reasoning Gap. Contrary to the expectation that debate sharpens reasoning [8], we found that enforcing a devil's advocate role introduced artificial doubt, causing the system to reject correct, high-probability diagnoses. This degradation was most severe in 'easier' diagnostic categories, where the mandate to critique over-complicated clear clinical signals. This degradation was most severe in easier diagnostic categories (e.g., Allergic, Rheumatological), where the debate process over-complicated clear clinical signals.
- We presented a domain-specific analysis which revealed that no single topology is universally superior. While the Hierarchical model has the highest diagnostic accuracy overall, the Collaborative approach demonstrated specific strengths in multi-organ pathologies, such as Respiratory and Urogenital diseases, where the synthesis of distinct diagnostic viewpoints (radiology, internal medicine, and pathology) is critical. Conversely, for highly ambiguous categories like Cardiac Malformation, all topologies struggled equally, indicating that the primary bottleneck in these instances is the opacity of the case data rather than the agent structure.

Ultimately, these findings demonstrate that increasing system complexity does not guarantee better reasoning. AI deployments should look beyond static architectures toward dynamic topology selection, where a supervisor system assigns a specific workflow e.g., utilizing a collaborative topology for multi-organ cases vs. a Single Agent for routine presentations [13].

## Dataset Availability

The dataset used in this study is provided by Chen et al. [9] and can be accessed at: https://github.com/geteff1/Multi-agent-conversation-for-disease-diagnosis